\documentstyle[12pt]{article}
\input epsf
\begin{document}

\title{Vector cross product in n-dimensional vector space}
\author{Xiu-Lao Tian$^{1}$\footnote{Corresponding author: Phone 86-29-8816-6089;
\ E-mail txl@xiyou.edu.cn;\ yang $_-$chaomail@163.com},
Chao Yang$^{1}$,~Yang Hu$^{1}$, ~Chao Tian$^{2}$ \\
\small {$^{1}$School of Science, Xi'an University of Posts and Telecommunications, Xi'an 710121, China}\\
\small {$^{2}$Engineering Department,Google Inc., Mountain View, CA 94043, USA}}

\maketitle
\begin{abstract}
The definition of vector cross product (VCP) introduced by Eckmann
only exists in thethree- and the seven- dimensional vector space. In
this paper, according to the orthogonal completeness, magnitude of
basis vector cross product and all kinds of combinations of basis
vector $\hat{e}_i$, the generalized definition of VCP in the odd
n-dimensional vector space is given by introducing a cross term
$X_{AB}$. In addition, the definition is validated by reducing the
generalization definition to the fundamental three- and
seven-dimensional vector space.
\end{abstract}


\maketitle
\section{Introduction}
Vector is an important mathematical tool, commonly used in physical
field, such as vector inner product, vector cross product (VCP), and
vector sensor product. It is well known that vector inner product
and tensor product can be existed in n-dimensional Euclidian space
(vector space). But for the VCP, it only exists in three- and
seven-dimensional vector space [1-4]. The VCP has wide application
in physics fields in three-dimensional space, e.g, torque, angular
momentum and so on, and in seven-dimensional VCP have been applied
at self-dual Yang-Mills fields [5-6] and Supergravity research [7],
etc. However, the VCP defined by Eckmann[1] is severely restricted
except the three- and seven-dimensional VCP. Consequently, the
extension of the definition of VCP to n-dimensional space has
important physical significance.

In this paper, a general definition of the VCP of two vectors on odd
n-dimensional space is given by employing the method of combination
in terms of orthogonal completeness and magnitude of basis vector
cross product and all kinds of combinations of basis vector
$\hat{e}_i$. The VCP $\vec{A}\times\vec{B}$ in odd n-dimensional
space satisfies the following two conditions:
(i)$(\vec{A}\times\vec{B},~ \vec{A}$ or $ \vec{B})=0$;

(ii)$\parallel\vec{A}\times\vec{B}\parallel^2=
\|\vec{A}\|^2\|\vec{B}\|^2-(\vec{A},\vec{B})^2 +X_{AB}$. When
$X_{AB}=0 $, the generalized definition of VCP corresponds to the
three- and seven-dimensional VCP defined by Eckmann [1].

\section{The generalized definition of the VCP in odd n-dimensional vector space}

Let V denote an n-dimensional vector space over the real numbers and
(,) denote the ordinary (positive definite) vector inner product.
The VCP $\vec{A}\times\vec{B}$ of any two vectors on V has been
defined by B. Eckmann [1] satisfying the following axioms

\begin{eqnarray}
(\vec{A}\times\vec{B}, \vec A ~or~ \vec B) =0, \\
 \parallel\vec{A}\times\vec{B}\parallel^2=
\|\vec{A}\|^2\|\vec{B}\|^2-(\vec{A},\vec{B})^2
\end{eqnarray}

Considering arbitrary two vectors $\vec A=a^i\hat{e}_i $ and $\vec
B=b^j\hat{e}_j$ in n-dimensional vector space,$\hat{e}_i$ and
$\hat{e}_j $ are basis vectors from the given orthogonal coordinate
system, $a^i $ and $ b^j $ are vector components corresponding to
$\vec A$ and $\vec B$, and the Einstein summation convention are
adopted.

The VCP $ \vec{A}\times\vec{B}$ can be expressed as
\begin{eqnarray}
\vec{A}\times\vec{B}=a^i\hat{e}_i\times
b^j\hat{e}_j=a^ib^j\hat{e}_i\times\hat{e}_j
\end{eqnarray}

Obviously, the magnitude of cross product is determined by these
vector components $a^i, b^j $ and the direction is determined by
basis vectors $\hat{e}_i\times\hat{e}_j $.

In the following, a generalized definition of VCP is presented based
on the orthogonal completeness, magnitude of VCP and all kinds of
combinations of basis vector.

Firstly, orthogonal completeness of the VCP requires the VCP only
exist in an odd n-dimensional space.

As $(\vec{A}\times\vec{B}, \vec A ~or~ \vec B) =0$, the cross
product of any two vectors is always perpendicular to both of the
vectors being multiplied and a plane containing them ( orthogonality
of the VCP ) and $\hat{e}_i\times\hat{e}_j$ of any two basis vectors
must be equal to another basis vector $ \hat{e}_k $,  i.e,
$\hat{e}_i\times\hat{e}_j=\pm\hat{e}_k$ ( completeness of the VCP).

Based on the definition of $\hat{e}_i\times\hat{e}_j= \hat{e}_k$, a
cross product $\hat{e}_i\times\hat{e}_j$ of any two basis vectors is
equivalent to a 2-combination of two basis vectors $\hat{e}_i$ and
$\hat{e}_j$. There are n basis vectors in n-dimensional vector
space, the number of 2-combination is that the number of combination
of n basis vectors taken 2 vectors at a time without repetitions.
The number of 2-combination of arbitrary two basis vectors in an
n-dimensional vector space is $ C_n^2=\frac{1}{{2}}n(n-1)$.

Taking the equality of each basis vector into account, so $ C_n^2 $
should be averagely distributed to each one basis vector, let K (
arbitrary integer ) denote the number of 2-combination which is
averagely distributed to each one basis vector. The number K is
determined by
 \begin{eqnarray}
K=\frac{C_n^2}{n}=\frac{1}{{2}}(n-1)
\end{eqnarray}
then
\begin{eqnarray}
n=2K+1.
\end{eqnarray}

So the VCP of two vectors there only exist in an odd n-dimensional
space.

Secondly, the definition of magnitude of the VCP in Eq.(2) can be generalized to
\begin{eqnarray}
\parallel\vec{A}\times\vec{B}\parallel^2=
\|\vec{A}\|^2\|\vec{B}\|^2-(\vec{A},\vec{B})^2 + X_{AB}
\end{eqnarray}
where $X_{AB}$ is called cross item and it can be expressed as
\begin{eqnarray}
X_{AB}=a_{i}b_{j}a^{l}b^{m}\chi^{ij}_{lm}=
a_{i}b_{j}a^{l}b^{m}[T^{ij}_{lm}+
\delta^i_m\delta^j_l-\delta^j_m\delta^i_l]
\end{eqnarray}
\begin{eqnarray}
T^{lm}_{ij}=
((\hat{e}_i\times\hat{e}_j)\cdot\hat{e}_k)
((\hat{e}_l\times\hat{e}_m)\cdot\hat{e}_k)
\end{eqnarray}
where $T^{lm}_{ij} $ is a sign function and
$((\hat{e}_i\times\hat{e}_j)\cdot\hat{e}_k) $ denote vector inner
product of $(\hat{e}_i\times\hat{e}_j)$ and $\hat{e}_k$.

Subsequently, the generalized definition and calculation formula of the VCP in
an odd n-dimensional space will be presented.

{\bf Proposition 1:} the VCP $\vec{A}\times\vec{B}$
 of any two vectors on an odd n-dimensional space satisfy the following generalized axioms:
\begin{eqnarray}
(i)~~~~~~~~~~~~~~~~~~~~~~~~~~~~ (\vec{A}\times\vec{B},~ \vec{A}~~  or ~~ \vec{B})=0\\
(ii)~~\parallel\vec{A}\times\vec{B}\parallel^2=
\|\vec{A}\|^2\|\vec{B}\|^2-(\vec{A},\vec{B})^2 +X_{AB}.
\end{eqnarray}

{\bf Proposition 2:} The calculation of any two VCP
$\vec{A}\times\vec{B}$ can be expressed by tensor

\begin{eqnarray}
\vec{A}\times\vec{B}=a^ib^j(\vec{e}_i\times\vec{e}_j)=a^ib^j
L_{ij~k}\hat{e}^k
\end{eqnarray}
\begin{eqnarray}
L_{ij~k}=(\vec{e}_i\times\vec{e}_j)\cdot\hat{e}^k
\end{eqnarray}
where  $L_{ij~k}=( \pm1, 0 )$ is a sign function (generalized
Levi-Civita symbol) on n-dimensional space, which is determined by a
fixed algorithm of cross product of basis vector, the algorithm will
be discussed in section 3.

\section{Algorithm of VCP in an odd n-dimensional space}
As you know, there is only an algorithm of VCP in 3-dimensional
space. Although the definition of the VCP has been extended to an
odd n-dimensional space $( n>3 )$, the algorithm of the VCP is not
unique. Owing to the  diversity of the combination of basis vector,
there are many kinds of algorithm of the VCP in odd n-dimensional
space. So-called an algorithm depend on a calculation rule. In the
following, it will give a detailed statement about the algorithm of
the VCP in an odd n-dimensional space.

\subsection{ Algorithm of the VCP in 5-dimensional vector space}
Obviously, there are 5 basis vectors ($\hat{e}_1, \hat{e}_2,
\hat{e}_3,\hat{e}_4,\hat{e}_5$) in 5-dimensional vector space , one
of the basis vectors can be expressed by the cross product of the
other two basis vector. From Eq. (4), one can find $k=2$, one of the
basis vectors can be expressed by two 2-combination of basis vector.
As is demonstrated in Table 1, e.g., the basis vector ~$\hat{e}_1$
can be expressed by $\hat{e}_2\times\hat{e}_3 ~( 23 )$ and  $
\hat{e}_4\times\hat{e}_5 ~( 45 )$. Here, a double-digit is used to
denote the cross product of two basis vectors for simplicity. The $(
23, 45 )$ of two 2-combination of basis vector is called one kind of
distributive combination form, there are 3 kinds of different
distributive combination forms under each basis vector $\hat{e}_i$.
For example, there are 3 kinds of distributive combination forms $(
23, 45 )$, $( 24, 35 )$ and $( 25, 34 )$ under $\hat{e}_1$ in Table
1.

\begin{center}
\centerline{\footnotesize Table 1. Three kinds of distributive
combination forms of basis vector under  $\hat{e}_i$ in
5-dimensional space}

\begin{tabular}{c|c|c|c|c|c}
\hline kind& $\hat e_1$ & $\hat e_2$ & $\hat e_3$ & $\hat e_4$ &
$\hat e_5$\cr \hline

1&23,~~45&15,~~34&12,~~45&13,~~25&14,~~23\\
&$\hat{e}_2\times\hat{e}_3$,$\hat{e}_4\times\hat{e}_5$
&$\hat{e}_5\times\hat{e}_1$,$\hat{e}_3\times\hat{e}_4$
&$\hat{e}_1\times\hat{e}_2$,$\hat{e}_5\times\hat{e}_4$
&$\hat{e}_1\times\hat{e}_3$,$\hat{e}_5\times\hat{e}_2$
&$\hat{e}_1\times\hat{e}_4$,$\hat{e}_2\times\hat{e}_3$\cr \hline

2&24,~~35&14,~~35&14,~~25&12,~~35&12,~~34\cr \hline

3&25,~~34&13,~~45&15,~~24&15,~~23&13,~~24\cr \hline
\end{tabular}
\end{center}

\begin{center} \centerline{\footnotesize Table 2. Six kinds of algorithm of
cross product in 5-dimensional space}
\begin{tabular}{c|c|c|c|c|c}
\hline algorithm &$\hat e_1$ & $\hat e_2$ & $\hat e_3$ & $\hat e_4$
& $\hat e_5$\cr \hline
1& 23 45 & 14 35 & 15 24 & 13 25 & 12 34 \cr \hline

2& 23 45 & 15 34 & 14 25 & 12 35 & 13 24 \cr \hline

{\bf3}& {\bf24 35} &{\bf 13 45 }&{\bf 14 25 }&{\bf 15 23 }&{\bf 12
34} \cr \hline

4& 24 35 & 15 34 & 12 45 & 13 25 & 14 23 \cr \hline

5& 25 34 & 13 45 & 15 24 & 12 35 & 14 23 \cr \hline

6& 25 34 & 14 35 & 12 45 & 15 23 & 13 24 \cr \hline

\end{tabular}
\end{center}

Furthermore, a pair of double-digit (a distributive combination
forms) from each column in Table 1 is taken to constitute a kind of
calculation rule, and the calculation rule demands all double-digit
of the five pair of double-digit which be extracted is different.
Then, let these different double-digit arrange a row to represent a
calculation rule. Accordingly, six kinds of different calculation
rules of basis vector are obtained in Table 2, that is to say, the
VCP in five-dimensional space have six sorts of different
algorithms. Thus, One can define a kind of cross product algorithm
of basis vector by selecting combination form of a row double-digit
from Table 2. For example, the relation of cross product of basis
vector from 3th row in Table 2 can be expressed as
\begin{eqnarray}
\hat{e}_2\times\hat{e}_4=\hat{e}_1,~\hat{e}_3\times\hat{e}_5=\hat{e}_1,~
\hat{e}_3\times\hat{e}_1=\hat{e}_2,~\hat{e}_4\times\hat{e}_5=\hat{e}_2,\nonumber\\
\hat{e}_4\times\hat{e}_1=\hat{e}_3,~\hat{e}_5\times\hat{e}_2=\hat{e}_3,~
\hat{e}_5\times\hat{e}_1=\hat{e}_4,~\hat{e}_2\times\hat{e}_3=\hat{e}_4,\nonumber\\
\hat{e}_1\times\hat{e}_2=\hat{e}_5,~\hat{e}_3\times\hat{e}_4=\hat{e}_5.
\end{eqnarray}

The above relation of cross product of basis vector shows a kind of
fixed algorithm of the VCP. Under right-handed orthogonal coordinate
frame ( or left-handed orthogonal coordinate frame ), the cross
product of basis vector have to satisfy right-handed rotation rule.
So the double-digit in Table 2 must be written into cross product of
two basis vector, e.g., ``13'' under $\hat{e}_2 $ in Table 2 must be
written as $\hat{e}_3\times\hat{e}_1=\hat{e}_2$, or
$\hat{e}_1\times\hat{e}_3=-\hat{e}_2$. Then, the relation of cross
product of basis vector can be expressed as
\begin{eqnarray}
\hat{e}_i\times\hat{e}_j=L_{ij~k}\hat{e}_k, ~~~~where~ repeated~index ~k ~does ~not ~sum
\end{eqnarray}
When $(ij~k)$ is even permutation, $L_{ij~k}=1$, when $(ij~k)$ is
odd permutation, $L_{ij~k}=-1$, when $(ij~k)$ is else permutation,
$L_{ij~k}=0$. Intuitively, a simple graphical expression of the
right-handed rotation is given in Fig. 1. to determine the sign of $
L_{ij~k} $.
\begin{center}
{\epsfxsize 100mm \epsfysize 35mm \epsffile{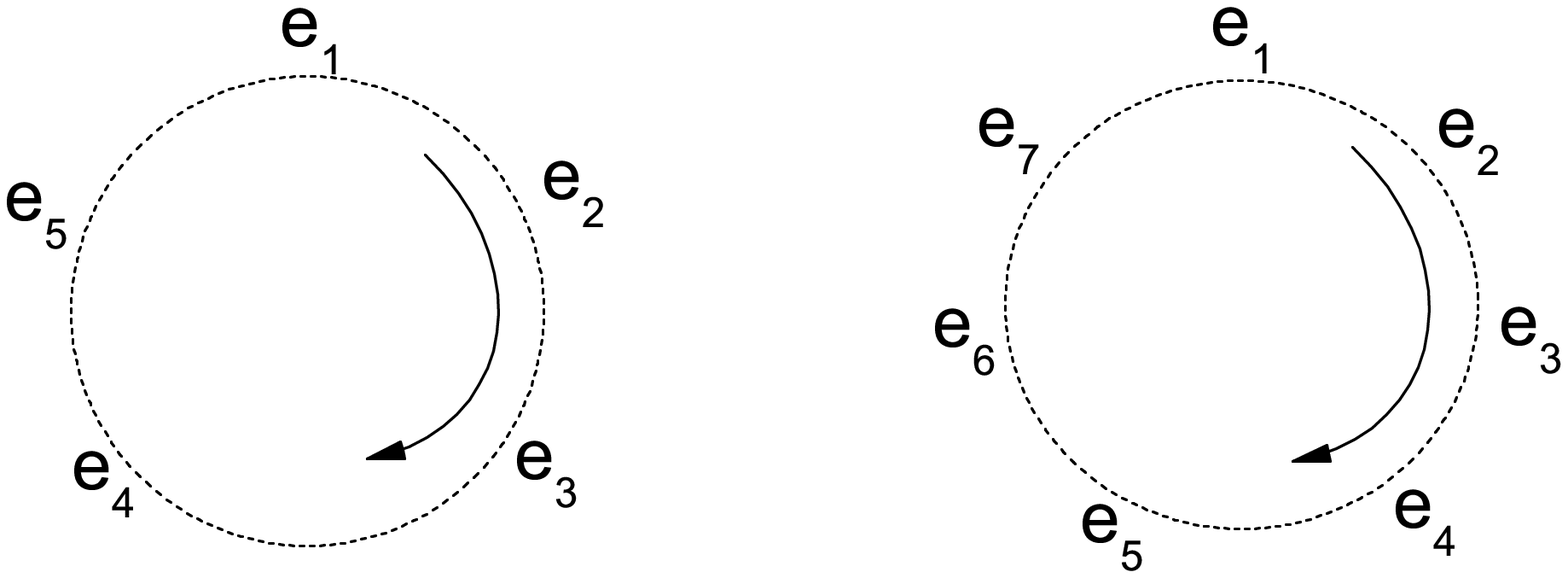}}

{\footnotesize Fig.1 right-handed rotation rule of cross product of
basis vectors}

\end{center}

In terms of the relation of basis vectors from Eq. (14), the VCP
$\vec{A}\times\vec{B}$ of any two vectors $\vec{A}$ and $\vec{B}$
can be computed by the following
\begin{eqnarray}
\vec{A}\times\vec{B}&=&a^ib^j(\vec{e}_i\times\vec{e}_j)=a^ib^j
L_{ijk}\hat{e}^k=a^ib^j\hat{e}^k
(\vec{e}_i\times\vec{e}_j)\cdot\hat{e}^k\nonumber\\
&=&\left[\left|\begin{array}{cc}
 a_2&a_4\\
 b_2&b_4
\end{array}\right|+\left|\begin{array}{cc}
 a_3&a_5\\
 b_3&b_5
\end{array}\right|\right
]\hat{e}_1+\left[\left|\begin{array}{cc}
 a_3&a_1\\
 b_3&b_1
\end{array}\right|+\left|\begin{array}{cc}
 a_4&a_5\\
 b_4&b_5
\end{array}\right|\right ]\hat{e}_2\nonumber\\
&+&\left[\left|\begin{array}{cc}
 a_4&a_1\\
 b_4&b_1
\end{array}\right|+\left|\begin{array}{cc}
 a_5&a_2\\
 b_5&b_2
\end{array}\right|\right]\hat{e}_3
+\left[\left|\begin{array}{cc}
 a_2&a_3\\
 b_2&b_3
\end{array}\right|+\left|\begin{array}{cc}
 a_5&a_1\\
 b_5&b_1
\end{array}\right|\right]\hat{e}_4\nonumber\\
&+&\left[\left|\begin{array}{cc}
 a_1&a_2\\
 b_1&b_2
\end{array}\right|+\left|\begin{array}{cc}
 a_3&a_4\\
 b_3&b_4
\end{array}\right|\right]\hat{e}_5
\end{eqnarray}
let $ X_{\alpha\beta}=\left|\begin{array}{cc}
 a_\alpha&a_\beta\\
 b_\alpha&b_\beta\end{array}\right|$ (determinant),  then $\vec{A}\times\vec{B}$
 can be expressed as
\begin{eqnarray}
\vec{A}\times\vec{B}=[X_{24}+X_{35}]\hat{e}_1+[X_{31}+X_{45}]\hat{e}_2+[X_{41}+X_{52}]\hat{e}_3\nonumber\\
+[X_{23}+X_{51}]\hat{e}_4+[X_{12}+X_{34}]\hat{e}_5
\end{eqnarray}

Moreover, we can also prove
$\parallel\vec{A}\times\vec{B}\parallel^2=
\|\vec{A}\|^2\|\vec{B}\|^2-(\vec{A},\vec{B})^2 +X_{AB}$. where
$X_{AB}=a^{i}b^{j}a_{l}b_{m}\chi^{lm}_{ij}$ is as follows
\begin{eqnarray}
X_{AB}=a^{i}b^{j}a_{l}b_{m}\chi^{lm}_{ij}=a^{i}b^{j}a_{l}b_{m}[T^{lm}_{ij}
+\delta^m_i\delta^l_j-\delta^m_j\delta^l_i]\nonumber\\
=a^{i}b^{j}a_{l}b_{m}[(\hat{e}_i
\times\hat{e}_j)\cdot\hat{e}_1(\hat{e}^l\times\hat{e}^m)\cdot\hat{e}^1
+\delta^m_i\delta^l_j-\delta^m_j\delta^l_i]\nonumber\\
+(\hat{e}_i
\times\hat{e}_j)\cdot\hat{e}_2(\hat{e}^l\times\hat{e}^m)\cdot\hat{e}^2
+\delta^m_i\delta^l_j-\delta^m_j\delta^l_i]\nonumber\\
+(\hat{e}_i
\times\hat{e}_j)\cdot\hat{e}_3(\hat{e}^l\times\hat{e}^m)\cdot\hat{e}^3
+\delta^m_i\delta^l_j-\delta^m_j\delta^l_i]\nonumber\\
+(\hat{e}_i
\times\hat{e}_j)\cdot\hat{e}_4(\hat{e}^l\times\hat{e}^m)\cdot\hat{e}^4
+\delta^m_i\delta^l_j-\delta^m_j\delta^l_i]\nonumber\\
+(\hat{e}_i
\times\hat{e}_j)\cdot\hat{e}_5(\hat{e}^l\times\hat{e}^m)\cdot\hat{e}^5
+\delta^m_i\delta^l_j-\delta^m_j\delta^l_i]\nonumber\\
=2[(a_2b_4-b_2a_4)(a_3b_5-b_3a_5)+(a_3b_1-b_3a_1)(a_4b_5-b_4a_5)\nonumber\\
+(a_4b_1-b_4a_1)(a_5b_2-b_5a_2)+(a_2b_3-b_2a_3)(a_5b_1-b_5a_1)\nonumber\\
+(a_1b_2-b_1a_2)(a_3b_4-b_3a_4)]\nonumber\\
 =2[X_{24}X_{35}+X_{31}X_{45}+X_{41}X_{52}
 +X_{23}X_{51}+X_{12}X_{34}]\neq0
\end{eqnarray}
\subsection{ Algorithm of the VCP in 7-dimensional vector space}

Similarly, we can find $k=3$ in 7-dimensional vector space using Eq.
(4). As is demonstrated in Table 3, each of basis vector $e_i$
includes three 2-combination of basis vector and there are 15 kinds
of different distributive combination forms.

\begin{table}

\begin{center}
\centerline{\footnotesize Table 3.  Fifteen kinds of distributive
combinatorial forms of basis vector under $\hat{e}_i$ in
7-dimensional space}
\begin{tabular}{c|c|c|c|c|c|c|c}
\hline kind &$\hat e_1$ & $\hat e_2$ & $\hat e_3$ & $\hat e_4$ &
$\hat e_5$ & $\hat e_6$ & $\hat e_7$ \cr \hline
1&24,37,56&14,35,67&17,25,46&12,36,57&16,23,47&15,27,34&13,26,45\cr
\hline
2&24,35,67&14,36,57&17,24,56&12,35,67&16,24,37&15,23,47&13,25,46\cr
\hline
3&24,36,57&14,37,56&17,26,45&12,37,56&16,27,34&15,24,37&13,24,56\cr
\hline
4&23,45,67&13,45,67&16,24,56&13,26,57&17,24,36&17,24,35&14,25,36\cr
\hline
5&23,46,57&13,46,57&16,25,47&13,25,67&17,23,46&17,23,45&14,23,56\cr
\hline
6&23,47,56&13,56,47&16,27,45&13,27,56&17,26,34&17,25,34&14,26,35\cr
\hline
7&25,37,46&15,34,67&15,24,67&15,26,37&14,26,37&14,25,37&15,26,34\cr
\hline
8&25,34,67&15,36,47&15,26,47&15,23,67&14,23,67&14,23,57&15,23,46\cr
\hline
9&25,36,47&15,37,46&15,27,46&15,27,36&14,27,36&14,27,35&15,24,36\cr
\hline
10&26,34,57&16,34,57&14,25,67&16,25,37&13,24,67&13,24,57&16,24,35\cr
\hline
11&26,35,47&16,35,47&14,26,57&16,23,57&13,26,47&13,25,47&16,23,45\cr
\hline
12&26,37,46&16,37,45&14,27,56&16,27,35&13,27,46&13,27,45&16,25,34\cr
\hline
13&27,34,56&17,34,56&12,45,67&17,25,36&12,34,67&12,34,57&12,34,56\cr
\hline
14&27,35,46&17,35,46&12,46,57&17,23,56&12,36,47&12,35,47&12,35,46\cr
\hline
15&27,36,45&17,36,45&12,47,56&17,26,35&12,37,46&12,37,45&12,36,45\cr
\hline
\end{tabular}
\end{center}
\end{table}

\begin{table}
\begin{center} \centerline{\footnotesize Table 4. Thirty kinds of algorithms
of cross product in 7-dimensional space}
\begin{tabular}{c|c|c|c|c|c|c|c}
\hline algorithm &$\hat e_1$ & $\hat e_2$ & $\hat e_3$ & $\hat e_4$
& $\hat e_5$&$\hat e_6$&$\hat e_7$\cr \hline

01&23 45 67 & 13 46 57 & 12 47 56 & 15 26 37 & 14 27 36 & 17 24 35 &16 25 34\cr \hline

02& 23 45 67 & 13 47 56 & 12 46 57 & 15 27 36 & 14 26 37 & 17 25 34& 16 24 35\cr \hline

03&23 46 57 & 13 45 67 & 12 47 56 & 16 25 37 & 17 24 36 & 14 27 35 &15 26 34 \cr \hline

04&23 46 57 & 13 47 56 & 12 45 67 & 16 27 35 & 17 26 34 & 14 25 37 &15 24 36\cr \hline

05& 23 47 56 & 13 45 67 & 12 46 57 & 17 25 36 & 16 24 37 & 15 27 34& 14 26 35\cr \hline

06& 23 47 56 & 13 46 57 & 12 45 67 & 17 26 35 & 16 27 34 & 15 24 37& 14 25 36\cr \hline

07& 24 35 67 & 14 36 57 & 15 26 47 & 12 37 56 & 13 27 46 & 17 23 45& 16 25 34\cr \hline

08&24 35 67 & 14 37 56 & 15 27 46 & 12 36 57 & 13 26 47 & 17 25 34 &16 23 45\cr \hline

09& 24 36 57 & 14 35 67 & 16 25 47 & 12 37 56 & 17 23 46 & 13 27 45& 15 26 34\cr \hline

10& 24 36 57 & 14 37 56 & 16 27 45 & 12 35 67 & 17 26 34 & 13 25 47& 15 23 46\cr \hline

\bf{11}&\bf{ 24 37 56 }&\bf{14 35 67 }&\bf{17 25 46 }&\bf{12 36 57
}& \bf{16 23 47 }&\bf{15 27 34 }&\bf{ 13 26 45}\cr \hline

12& 24 37 56 & 14 36 57 & 17 26 45 & 12 35 67 & 16 27 34 & 15 23 47& 13 25 46\cr \hline

13& 25 34 67 & 15 36 47 & 14 26 57 & 13 27 56 & 12 37 46 & 17 23 45& 16 24 35\cr \hline

14& 25 34 67 & 15 37 46 & 14 27 56 & 13 26 57 & 12 36 47 & 17 24 35& 16 23 45\cr \hline

15& 25 36 47 & 15 34 67 & 16 24 57 & 17 23 56 & 12 37 46 & 13 27 45& 14 26 35\cr \hline

16& 25 36 47 & 15 37 46 & 16 27 45 & 17 26 35 & 12 34 67 & 13 24 57& 14 23 56\cr \hline

17& 25 37 46 & 15 34 67 & 17 24 56 & 16 23 57 & 12 36 47 & 14 27 35& 13 26 45\cr \hline

18& 25 37 46 & 15 36 47 & 17 26 45 & 16 27 35 & 12 34 67 & 14 23 57& 13 24 56\cr \hline

19& 26 34 57 & 16 35 47 & 14 25 67 & 13 27 56 & 17 23 46 & 12 37 45& 15 24 36\cr \hline

\bf{20}&\bf{ 26 34 57 }& \bf{16 37 45} & \bf{14 27 56} & \bf{13 25 67 }& \bf{17 24 36} & \bf{12 35 47} & \bf{15 23 46}\cr \hline

21& 26 35 47 & 16 34 57 & 15 24 67 & 17 23 56 & 13 27 46 & 12 37 45& 14 25 36\cr \hline

22& 26 35 47 & 16 37 45 & 15 27 46 & 17 25 36 & 13 24 67 & 12 34 57& 14 23 56\cr \hline

23& 26 37 45 & 16 34 57 & 17 24 56 & 15 23 67 & 14 27 36 & 12 35 47& 13 25 46\cr \hline

24& 26 37 45 & 16 35 47 & 17 25 46 & 15 27 36 & 14 23 67 & 12 34 57& 13 24 56\cr \hline

25& 27 34 56 & 17 35 46 & 14 25 67 & 13 26 57 & 16 23 47 & 15 24 37& 12 36 45\cr \hline

26& 27 34 56 & 17 36 45 & 14 26 57 & 13 25 67 & 16 24 37 & 15 23 47& 12 35 46\cr \hline

27& 27 35 46 & 17 34 56 & 15 24 67 & 16 23 57 & 13 26 47 & 14 25 37& 12 36 45\cr \hline

28& 27 35 46 & 17 36 45 & 15 26 47 & 16 25 37 & 13 24 67 & 14 23 57& 12 34 56\cr \hline

29& 27 36 45 & 17 34 56 & 16 24 57 & 15 23 67 & 14 26 37 & 13 25 47& 12 35 46\cr \hline

30& 27 36 45 & 17 35 46 & 16 25 47 & 15 26 37 & 14 23 67 & 13 24 57& 12 34 56\cr \hline
\end{tabular}
\end{center}
\end{table}
One can take a combinatorial forms which include 3 double-digit from
each column that include 15 kinds of combinatorial forms under each
basis vector $\hat e_i$, and these 7 kinds of different 3
double-digit also arrange a row. By means of computer, there are
6240 kinds of no repeating double-digit combinatorial forms. Namely,
there are 6240 kinds of algorithms of basis vector in 7-dimensional
vector space. Next, we show 30 kinds of no repeating double-digit
combinatorial forms in Table 4. From the basis vector of 11th row
and the 20th row in Table 4, it is just the previous rule of VCP in
7-dimensional vector space defined by B.Eckmann.

For the algorithm of the 11th row in Table 4. the relation of cross
product of basis vectors are expressed as

\begin{eqnarray}
\hat{e}_2\times\hat{e}_4=\hat{e}_1,~~\hat{e}_3\times\hat{e}_7=\hat{e}_1,~~
\hat{e}_5\times\hat{e}_6=\hat{e}_1,\nonumber\\
\hat{e}_4\times\hat{e}_1=\hat{e}_2,~~\hat{e}_3\times\hat{e}_5=\hat{e}_2,~~
\hat{e}_6\times\hat{e}_7=\hat{e}_2,\nonumber\\
\hat{e}_4\times\hat{e}_6=\hat{e}_3,~~\hat{e}_5\times\hat{e}_2=\hat{e}_3,~~
\hat{e}_7\times\hat{e}_1=\hat{e}_3,\nonumber\\
\hat{e}_5\times\hat{e}_7=\hat{e}_4,~~\hat{e}_6\times\hat{e}_3=\hat{e}_4,~~
\hat{e}_1\times\hat{e}_2=\hat{e}_4,\nonumber\\
\hat{e}_6\times\hat{e}_1=\hat{e}_5,~~\hat{e}_7\times\hat{e}_4=\hat{e}_5,~~
\hat{e}_2\times\hat{e}_3=\hat{e}_5,\nonumber\\
\hat{e}_7\times\hat{e}_2=\hat{e}_6,~~\hat{e}_1\times\hat{e}_5=\hat{e}_6,~~
\hat{e}_3\times\hat{e}_4=\hat{e}_6,\nonumber\\
\hat{e}_1\times\hat{e}_3=\hat{e}_7,~~\hat{e}_2\times\hat{e}_6=\hat{e}_7,~~
\hat{e}_4\times\hat{e}_5=\hat{e}_7.
\end{eqnarray}

For any two vectors  $\vec A=a^i\hat{e}_i$ and $\vec B=b^j\hat{e}_j$
in 7-dimensional vector space,  the $\vec{A}\times\vec{B}$ can be expressed as
\begin{eqnarray}
\vec{A}\times\vec{B}=[X_{24}+X_{37}+X_{56}]\hat{e}_1+[X_{41}+X_{35}+X_{67}]\hat{e}_2+[X_{71}+X_{52}+X_{46}]\hat{e}_3\nonumber\\
+[X_{12}+X_{63}+X_{57}]\hat{e}_4+[X_{61}+X_{23}+X_{74}]\hat{e}_5+[X_{15}+X_{72}+X_{34}]\hat{e}_6\nonumber\\
+[X_{13}+X_{26}+X_{45}]\hat{e}_7
\end{eqnarray}

Certainly, one can verify

\begin{eqnarray}X_{AB}=2a^{i}b^{j}a_{l}b_{m}T^{lm}_{ij}=2\{[X_{24}X_{37}+
X_{24}X_{56}+X_{37}X_{56}]+[X_{41}X_{35}+X_{41}X_{67}+X_{35}X_{67}]\nonumber\\
+[X_{71}X_{52}+X_{71}X_{46}+X_{52}X_{46}]+[X_{12}X_{63}+X_{12}X_{57}+X_{63}X_{57}]\nonumber\\
+[X_{61}X_{23}+X_{61}X_{74}+X_{23}X_{74}]+[X_{15}X_{72}+X_{15}X_{34}+X_{72}X_{34}]\nonumber\\
+[X_{13}X_{26}+X_{13}X_{45}+X_{26}X_{45}]\}=0.
\end{eqnarray}
Therefore, the magnitude of vector cross product satisfy axiom
 $\parallel\vec{A}\times\vec{B}\parallel^2=
\|\vec{A}\|^2\|\vec{B}\|^2-(\vec{A},\vec{B})^2$,  i.e.,
$X_{AB}=a^{i}b^{j}a_{l}b_{m}T^{lm}_{ij}=0$

For the algorithm of the 20th row in Table 4. the relation of cross
product of basis vectors are expressed as
\begin{eqnarray}
\hat{e}_2\times\hat{e}_6=\hat{e}_1,~~\hat{e}_3\times\hat{e}_4=\hat{e}_1,~~
\hat{e}_5\times\hat{e}_7=\hat{e}_1,\nonumber\\
\hat{e}_6\times\hat{e}_1=\hat{e}_2,~~\hat{e}_3\times\hat{e}_7=\hat{e}_2,~~
\hat{e}_4\times\hat{e}_5=\hat{e}_2,\nonumber\\
\hat{e}_4\times\hat{e}_1=\hat{e}_3,~~\hat{e}_7\times\hat{e}_2=\hat{e}_3,~~
\hat{e}_5\times\hat{e}_6=\hat{e}_3,\nonumber\\
\hat{e}_1\times\hat{e}_3=\hat{e}_4,~~\hat{e}_5\times\hat{e}_2=\hat{e}_4,~~
\hat{e}_6\times\hat{e}_7=\hat{e}_4,\nonumber\\
\hat{e}_7\times\hat{e}_1=\hat{e}_5,~~\hat{e}_2\times\hat{e}_4=\hat{e}_5,~~
\hat{e}_6\times\hat{e}_3=\hat{e}_5,\nonumber\\
\hat{e}_1\times\hat{e}_2=\hat{e}_6,~~\hat{e}_3\times\hat{e}_5=\hat{e}_6,~~
\hat{e}_3\times\hat{e}_4=\hat{e}_6,\nonumber\\
\hat{e}_1\times\hat{e}_5=\hat{e}_7,~~\hat{e}_2\times\hat{e}_3=\hat{e}_7,~~
\hat{e}_4\times\hat{e}_6=\hat{e}_7.
\end{eqnarray}

Similarly, for any two vectors $\vec A$ and $\vec B$, there exist
$\parallel\vec{A}\times\vec{B}\parallel^2=
\|\vec{A}\|^2\|\vec{B}\|^2-(\vec{A},\vec{B})^2$. i.e.,
$X_{AB}=a^{i}b^{j}a_{l}b_{m}T^{lm}_{ij}=0$

Importantly, we pay more attention to the case of $ X_{AB}\neq 0 $,
and we take the algorithm of the 2th row in Table 4 as an example,
the relation of cross product of basis vectors are expressed as
\begin{eqnarray}
\hat{e}_2\times\hat{e}_3=\hat{e}_1,~~\hat{e}_4\times\hat{e}_5=\hat{e}_1,~~
\hat{e}_6\times\hat{e}_7=\hat{e}_1,\nonumber\\
\hat{e}_3\times\hat{e}_1=\hat{e}_2,~~\hat{e}_4\times\hat{e}_7=\hat{e}_2,~~
\hat{e}_5\times\hat{e}_6=\hat{e}_2,\nonumber\\
\hat{e}_1\times\hat{e}_2=\hat{e}_3,~~\hat{e}_4\times\hat{e}_6=\hat{e}_3,~~
\hat{e}_5\times\hat{e}_7=\hat{e}_3,\nonumber\\
\hat{e}_5\times\hat{e}_1=\hat{e}_4,~~\hat{e}_7\times\hat{e}_2=\hat{e}_4,~~
\hat{e}_6\times\hat{e}_3=\hat{e}_4,\nonumber\\
\hat{e}_1\times\hat{e}_4=\hat{e}_5,~~\hat{e}_6\times\hat{e}_2=\hat{e}_5,~~
\hat{e}_7\times\hat{e}_3=\hat{e}_5,\nonumber\\
\hat{e}_7\times\hat{e}_1=\hat{e}_6,~~\hat{e}_2\times\hat{e}_5=\hat{e}_6,~~
\hat{e}_3\times\hat{e}_4=\hat{e}_6,\nonumber\\
\hat{e}_1\times\hat{e}_6=\hat{e}_7,~~\hat{e}_2\times\hat{e}_4=\hat{e}_7,~~
\hat{e}_3\times\hat{e}_5=\hat{e}_7.
\end{eqnarray}

For any two vectors  $\vec A$ and $\vec B$ in 7-dimensional space,  the
$\vec{A}\times\vec{B}$ can be expressed as
\begin{eqnarray}
\vec{A}\times\vec{B}=[X_{23}+X_{45}+X_{67}]\hat{e}_1+[X_{32}+
X_{46}+X_{57}]\hat{e}_2+[X_{12}+X_{47}+X_{56}]\hat{e}_3\nonumber\\
+[X_{51}+X_{46}+X_{73}]\hat{e}_4+[X_{14}+X_{57}+X_{63}]\hat{e}_5+
[X_{71}+X_{24}+X_{35}]\hat{e}_6\nonumber\\
+[X_{16}+X_{25}+X_{34}]\hat{e}_7
\end{eqnarray}

In addition, one can verify

\begin{eqnarray}X_{AB}=a^{i}b^{j}a_{l}b_{m}T^{lm}_{ij}=2\{[X_{23}X_{45}+
X_{23}X_{67}+X_{45}X_{67}]+[X_{31}X_{46}+X_{31}X_{57}+X_{46}X_{57}]\nonumber\\
+[X_{12}X_{47}+X_{12}X_{56}+X_{47}X_{56}]+[X_{51}X_{62}+X_{51}X_{73}+X_{62}X_{73}]\nonumber\\
+[X_{14}X_{72}+X_{14}X_{63}+X_{72}X_{63}]+[X_{71}X_{24}+X_{71}X_{35}+X_{24}X_{35}]\nonumber\\
+[X_{16}X_{25}+X_{16}X_{34}+X_{25}X_{34}]\}\neq 0
\end{eqnarray}

From the above discussion, a kind of algorithm of the VCP is
determined once the relation of cross product in an odd
n-dimensional vector is chosen. And, the VCP satisfies axiom
$\parallel\vec{A}\times\vec{B}\parallel^2=
\|\vec{A}\|^2\|\vec{B}\|^2-(\vec{A},\vec{B})^2 $ is only a special
instance of $\parallel\vec{A}\times\vec{B}\parallel^2=
\|\vec{A}\|^2\|\vec{B}\|^2-(\vec{A},\vec{B})^2+X_{AB} $ proposed in
this paper. Of course, different algorithms will yield different
results. Probably, different algorithms of the VCP correspond to
different physical problems.

\section{Conclusion}
In this paper, the definition of the VCP defined by Eckmann has been
extended to an odd n-dimensional space by introducing a cross term $
X_{AB} $, and the results show that the new generalized definition
can be reduced to Eckmann's definition in three- and
seven-dimensional vector space. It should be noted that the result
of VCP $ \vec{A}\times\vec{B} $ depend on the choice of relation of
basis vector cross product, and different algorithms may correspond
to different physical problems, this need to be studied in the
future. Its potential applications in quantum physics [8,9], also
deserves to be further investigated.

\section*{ACKNOWLEDGMENTS}
We thank prof.  X. Q. Xi, C. X. Li and  H. Z. Qu for helpful
discussions. This work was supported by NSFC (11205121, 10974247,
11175248), the NSF of Shaanxi Province (2010JM1011), and the
Scientific Research Program of Education Department of Shaanxi
Provincial Government (12JK0992,12JK0986).

\end{document}